\documentclass[12pt,a4paper]{article}

\usepackage{graphicx}
\usepackage{amsmath,amssymb,mathrsfs}
\usepackage{cite}
\usepackage{bm}

\usepackage{enumitem}
\usepackage{color}
\usepackage{hyperref}
\usepackage{enumerate}
\usepackage{verbatim}

\usepackage{setspace}
\date{\today}

\def\be{\begin{equation}}
\def\ee{\end{equation}}

\author{Aron C. Wall\footnote{aroncwall@gmail.com}
\\ \textit{\small School of Natural Sciences, Institute for Advanced Study}
\\ \textit{\small 1 Einstein Dr, Princeton NJ, 08540 USA} }
\title{A Second Law for Higher Curvature Gravity}

\DeclareSymbolFont{matha}{OML}{txmi}{m}{it}
\DeclareMathSymbol{v}{\mathord}{matha}{118}

\begin{document}

\maketitle


\begin{abstract}
The Second Law of black hole thermodynamics is shown to hold for arbitrarily complicated theories of higher curvature gravity, so long as we allow only linearized perturbations to stationary black holes.  Some ambiguities in Wald's Noether charge method are resolved.  The increasing quantity turns out to be the same as the holographic entanglement entropy calculated by Dong.  It is suggested that only the linearization of the higher-curvature Second Law is important, when consistently truncating a UV-complete quantum gravity theory.
\end{abstract}

You've just invented a new theory of gravity.  Like Einstein's, your theory is generally covariant and formulated in terms of a metric $g_{ab}$.  But the action is more complicated; it is an arbitrary function of the Riemann curvature tensor, perhaps some scalars $\phi$, and their derivatives:
\be
I = \int d^Dx\,\sqrt{-g}\left( L_g(g_{ab}, R_{abcd}, \nabla_a R_{bcde}, \nabla_a \nabla_b R_{cdef}\ldots \phi, \nabla_a \phi, \nabla_a \nabla_b \phi \ldots) + L_m \right)
\ee
where $L_g$ is the exciting gravitational piece, while $L_m$ is a boring minimally coupled matter sector obeying the null energy condition $T_{ab} k^a k^b \ge 0$ (NEC) ($k^a$ being null).  Such ``higher curvature'' corrections are known to occur with small coefficients due to quantum and/or stringy corrections.

The equation of motion from varying the metric is
\be\label{EOM}
H_{ab} \equiv \frac{2}{\sqrt{-g}}\frac{\delta L_g}{g^{ab}} = 
\frac{-2}{\sqrt{-g}}\frac{\delta L_m}{g^{ab}} \equiv T_{ab}.
\ee
To make up for all the excitement in the gravitational sector, you decide the matter sector should be an ordinary field theory minimally coupled to $g_{ab}$, obeying the null energy condition $T_{ab} k^a k^b \ge 0$ (NEC) for $k^a$ null.

A lesser mind would ask whether this theory is in agreement with observation, or perhaps whether the vacuum is even stable.  But not you!  You are concerned with a far deeper question: do black holes in your theory still obey the Second Law of horizon thermodynamics?	

In GR, the NEC (plus a version of cosmic censorship) implies that the area of any future event horizon $H$ is always increasing \cite{Areathm}.  So Bekenstein \cite{Bek} postulated that black holes have entropy proportional to their area; Hawking radiation \cite{Hawking} showed that it was more than just an analogy, and that in turn had all kinds of ramifications \cite{toomany}!  Does this only make sense for the Einstein-Hilbert action, or is it true more broadly?  We shall see that there is indeed a Second Law in all such theories of higher curvature gravity theories, provided that you only consider \emph{linearized} perturbations $\delta g_{ab}$, $\delta \phi$ of the gravitational fields (possibly sourced by a first order perturbation to $\delta T_{ab}$) evaluated on a stationary black hole background (or more precisely, a bifurcate Killing horizon\footnote{At least through Eq. \eqref{SJKM}, the arguments below can also be adapted to linearized perturbations of stationary null surfaces in pp-wave spacetimes, using the null translation symmetry instead of the Killing boost.}).  This has previously been done for $f(\text{Lovelock})$ gravity \cite{fLove}, and quadratic curvature gravity \cite{BSW}.

Pick a gauge so that $u$ and $v$ are null coordinates which increase as one moves spacelike away from the horizon, so that $u = 0$ is the future horizon $H$, $v$ is an affine parameter along the null generators of the horizon, $v = 0$ is the past horizon, $i,j$ indices point in the $D - 2$ transverse directions, the metric obeys
\begin{eqnarray}
g_{vv} = g_{vi} = \,\,\,\,\,&\!\!\!\!\! g_{vv,u} = 0,\,\,\,g_{uv} = 1 \qquad (u = 0);& \\
g_{uu} = g_{ui} = 0 & \qquad \qquad \qquad \,\,(\text{everywhere});&
\end{eqnarray}
and the Killing symmetry acts like a standard Lorentz boost on the null coordinates: $v \to av$, $u \to u/a$.\footnote{On a slice of the horizon, this gauge implies that the only nonvanishing Christoffel symbols are $\Gamma^{k}_{ij}$ (the intrinsic geometry), $\Gamma_{ij}^v$, $\Gamma_{iu}^j$, $\Gamma_{ij}^u$, $\Gamma_{iv}^j$ (which can be calculated from the extrinsic curvatures $K_{ij(u)}$ and $K_{ij(v)}$), and $\Gamma_{iv}^v$, $\Gamma_{iu}^u$, $\Gamma_{uv}^i$ (the twist).}

The Killing weight $n$ of a tensor (with all indices lowered) is given by the number of $v$-indices minus the number of $u$-indices, and will sometimes be indicated by an $^{(n)}$ superscript.  A key feature of this gauge choice is that on the horizon, any tensor with positive weight $n$ always has at least $n$ $v$-derivatives acting on it.\footnote{If your theory had any gravitational fields with spin besides the metric, it would probably be necessary to provide them with a gauge symmetry too, in order to permit an analogous gauge-fixing.  For example, in the case of a vector potential you would need to impose the null gauge $A_v = 0$.  But this would raise additional questions e.g. is the entropy always gauge-invariant even when $A_a$ is nonminimally coupled?  So for now, you reluctantly stick to a metric-scalar theory.}

Now the first question is \emph{which entropy} should you use, which might always increase?  It should be some local geometrical expression which can be integrated along a given horizon slice, but which one?

Unfortunately, the Noether charge method \cite{Wald,IW} used to derive the black hole entropy is subject to a number of ambiguities, identified by Jacobson, Kang, and Myers (JKM) \cite{JKM,IW}.  None of these ambiguities matter for compact, stationary horizons.  For compact but nonstationary horizons there are ambiguities of the form \cite{JKM,fLove}
\be
S^{(\text{JKM})} = \int d^{D-2}x\,\sqrt{g} \sum X^{(n)} \cdot Y^{(-n)},
\ee
where the expression is a boost invariant (i.e. Killing weight 0) product of terms which are not separately boost invariant.  Such ambiguity terms vanish on Killing horizons.  At first order they vanish on the bifurcation surface $v = 0$, but they may be important at $v \ne 0$.  Possible Noether charges include:
\begin{description}
  \item[Wald] Differentiate with respect to the Riemann tensor (integrating by parts where necessary) \cite{Wald,IW}:\vspace{-1ex}
\be
S_\text{Wald} = -2\pi \int d^{D-2}x\,\sqrt{g}\,4\,\frac{\delta L_g}{\delta R_{uvuv}}.
\ee
\vspace{-2.5ex}
  \item[Iyer-Wald] Expand the Wald entropy in fields and their derivatives, and keeping the terms which depend only on boost-invariant (weight 0) fields \cite{IW}.
  \item[Dong] The holographic entanglement entropy in $f(\text{Riemann})$ theories, derived by analytically continuing certain gravitational instantons \cite{Dong}.\footnote{Some special cases are given in \cite{S1,S2,S3,S4,S5,S6}, while progress for actions with derivatives of Riemann is in \cite{R1,R2,R3}.}
\end{description}

Every entropy in the class $S_\text{Wald} + S^{(\text{JKM})}$ obeys the physical process version of the First Law,\footnote{For higher curvature gravity theories, this was shown explicitly in \cite{JKM2}, and is implied by the Appendix of \cite{IW} or section 2 of \cite{GW}.} which says that if the horizon begins and ends in a stationary configuration, the increase in entropy from $v = 0$ to $v = +\infty$ due to a first order perturbation $\delta g_{ab}, \delta \phi$ is given by the Killing energy flux across the horizon.
\be\label{1st}
\delta S(+\infty) - \delta S(0) = 
\int_{v = 0}^{+\infty} \delta T_{vv}\, v\,dv\,\sqrt{g}\,d^{D-2}x.
\ee
Given the NEC, it follows that $\Delta S \ge 0$.  The JKM ambiguity doesn't matter because factors with negative weight vanish at $v = 0$ while factors with positive weight vanish as $v \to +\infty$.

So does this prove the Second Law?  Not yet, because the Second Law requires the entropy to be increasing at every instant of time: $dS/dv \ge 0$.  At intermediate times, you must fix the JKM ambiguity (up to higher order terms like $K^4$ which vanish at linear order).

The key is to rewrite $H_{vv} = T_{vv} \ge 0$ on the horizon as the second derivative of some quantity $\varsigma(g_{ab}, \delta g_{ab}, \phi, \delta \phi)$:
\be\label{2nd}
\delta \int d^{D-2}x \sqrt{g} H_{vv} = -2\pi \partial_v \partial_v \varsigma \ge 0.
\ee
You can show this by expanding $\delta (\sqrt{g} H_{vv})$ (which has weight 2) as a sum of product of the gravitational fields and their variations:
\be\label{XY}
\delta (\sqrt{g} H_{vv}) = \sum_{n \ge 0} X^{(-n)} \delta Y^{(2+n)},
\ee
where the positivity of $n$ arises due to the fact that all fields with positive weight vanish on a future Killing horizon (since otherwise the Killing symmetry would require it to diverge at the bifurcation surface), and $Y$ takes the form $g_{ab,\ldots}$ or $\phi_{,ab\ldots}$.  By virtue of the gauge fixing, the total number of $v$-derivatives must always be at least $2 + n$.  By repeatedly differentiating by parts, you can move each of these derivatives either (a) to the $X$ term or (b) outside the expression entirely.  But (a) can happen at most $n$ times since the $X$ term cannot have positive weight.  So at least two $\partial_v$'s end up outside the expression, proving \eqref{2nd}.

If the black hole becomes stationary at late times, you can impose the final boundary condition 
$\partial_v \varsigma(+\infty) = 0$.  (Since the Killing time is $t = e^v$, this condition requires only that the perturbation to the black hole grows slower than exponentially with respect to Killing time.)  By integrating backwards in time using \eqref{2nd}, you can therefore conclude that $\partial_v \varsigma \ge 0$, which looks very much like a Second Law.

However, you probably would also like to know that $\varsigma$ is exact, i.e. a variation of some local geometric entropy: $\varsigma = \delta S$.  To see this, expand $\varsigma$ as sum of a product of terms of the form
\be
\varsigma = \sum X^{(0)} \delta Y^{(0)} + \sum_{n \ge 1} X^{(-n)} \delta Y^{(n)}.
\ee
where the $X$ terms contain no factors with positive weight.  In the terms with $n \ge 1$, since on the background spacetime $Y = 0$ on $H$, $(\delta X)Y = 0$ and thus you can simply move the $\delta$ to the outside, obtaining a term of the JKM form:
\be\label{SJKM}
\delta \sum_{n \ge 1} X^{(-n)} Y^{(n)} = \delta S^{({JKM})}.
\ee

The term made of boost-invariant factors (call it $\varsigma_0$) is trickier, but using the physical process version of the First Law, you can see that it is the variation of the Iyer-Wald entropy $S_\text{IW}$.  Let $B$ be the space of all possible boost-invariant metric data for first order variations on a given slice with $v = \text{const}$, with arbitrary matter sources.  Considering only boost-invariant data you can't tell whether or not you are on the bifurcation surface $v = 0$, so without loss of generality let's suppose you are (for purposes of evaluating $\varsigma_0$).

Now if the horizon becomes stationary at late times, you can always choose $\delta g_{ab}$, $\delta \phi$ to vanish at late times on the horizon, so that for any $\delta g_{ab},\,\delta \phi \in B$,
\be
\varsigma_0 = -\int_{v = 0}^{+\infty} \delta T_{vv}\, v\,dv\,\sqrt{g}\,d^{D-2}x
= \delta S_\text{IW},
\ee
where the first equality comes from integrating \eqref{2nd}, and the second from \eqref{1st}.  You can thus conclude that there is an increasing Noether charge entropy, of the form
\be
S = S_\text{IW} + S^{(\text{JKM})},
\ee
where the JKM ambiguity term can be constructed explicitly by the procedure given above, using only those terms in $H_{vv}$ which have at least three $\partial_v$'s acting on the same field.\footnote{By replacing $v$ with a general affine coordinate $\lambda = a + bv$, $b > 0$, where $a, b$ can depend on the generator of the horizon, and by performing the same series of steps, it follows that $S$ is also increasing from any initial slice on the horizon to any final slice.  (Note that, in the step which uses the physical process First Law, the bifurcation surface is not generally at a constant value of $\lambda$, but this does not invalidate the proof.)  It is perhaps no longer totally obvious that the increasing entropy formula is a covariant functional of the horizon slice (i.e. that it does not depend on b), but a direct calculation for $f(\text{Riemann})$ shows that the entropy is covariant at least in this case.}

In the special case of $f(\text{Riemann})$ gravity, with some effort (see Appendix A if you get stuck) you can show that (up to a total derivative, which vanishes if the horizon is compact):
\be\label{Dong}
S = -2\pi \int d^{D-2}x\,\sqrt{g}\left[4\,\frac{\partial L_g}{\partial R_{uvuv}}
+ 16\,\frac{\partial^2 L_g}{\partial R_{uiuj} \partial R_{vkvl}} K_{ij(u)} K_{kl(v)} \right],
\ee
where $K_{ij(a)}$ is the extrinsic curvature in the $a$ direction.  This matches $S_\text{Dong}$ exactly at linear order in the metric perturbation!\footnote{Note that $\mathcal{O}(K^4) = K_{(u)}^2 K_{(v)}^2$ and higher order terms cannot be determined by the linearized second law \cite{fLove}.  These terms are given in \cite{Dong} by a more complicated expression involving a symmetry factor.  According to \cite{R1, R2}, there is a ``splitting problem'' which renders Dong's method ambiguous for terms of order $K^4$ and higher, but fortunately this problem does not matter at $\mathcal{O}(K^2)$.}  This is a rather remarkable coincidence, considering that the two entropies were calculated using completely different techniques.

If your gravitational theory is coupled semiclassically to a a free quantum field theory, a Generalized Second Law (which accounts for the entropy in Hawking radiation and matter fields) also follows, as in \cite{rapid,fLove}.

At second order in the metric perturbation, one could consider the effects of gravity waves on the Second Law.  But some theories (e.g. GR with a negative $G_N$, in which gravity waves carry negative energy) will fail this test.  But at very nonlinear order (e.g. black hole mergers), even Lovelock gravity violates the Second Law \cite{JM,Liko,merge}!  Perhaps this indicates that it is only consistent to treat the non-GR couplings perturbatively, in a consistent truncation of a UV-complete theory of quantum gravity \cite{juan,critique}.  But in adiabatic (i.e. thermodynamically reversible) quantum processes for which $\delta (S + S_{\text{matter}}) = 0$, the linearized Second Law could still be a necessary consistency condition, even in a perturbative treatment \cite{BSW}.

\small
\subsection*{Acknowledgments}
I am grateful for interactions with Sudipta Sarkar, Srijit Bhattacharjee, Ted Jacobson, Zach Fisher, and Will Kelly, and for support from NSF grants PHY-1314311, PHY11-25915, the Institute for Advanced Study, UC Berkeley and the KITP.
\normalsize

\appendix

\section{The entropy formula for f(Riemann)}

This appendix details the calculation of the increasing entropy $S = S_\text{IW} + S^{(\text{JKM})}$ for $f(\text{Riemann})$ gravity.  In a moment we will catalogue the terms that arise, but first we will comment on their structure.

Each term of $H_{vv}$ arises either from varying either an inverse metric $g^{ab}$ or a Riemann tensor $R_{abcd}$ with respect to $g^{vv}$.  Only the later process produces terms with derivatives of the Riemann tensor, which is needed to get a $S^{(\text{JKM})}$ term.

Thus we look for a Riemann term with the index structure $R_{uaub}$, and eliminate it while integrating by parts to move the $\nabla_a$ and $\nabla_b$ derivatives so that they act on the remaining parts of the expression, using the relation
\begin{eqnarray}\label{diffR}
\int d^Dx\,\sqrt{-g}\,X^{abcd} \delta R_{cadb} = 
-\frac{1}{2} \int d^Dx\,\sqrt{-g}\,\nabla_a \nabla_b X^{abcd} \delta g_{cd} \\ \nonumber
+ (a \leftrightarrow -c,\, b \leftrightarrow -d)\,\text{permutations} + \text{curvature\,terms},
\end{eqnarray}
where the curvature terms involve Riemann instead of derivatives of $X^{abcd}$.  This relation can be derived most easily in local inertial coordinates.

To obtain the JKM terms, we then look for terms with a factor of weight 3 or greater; these either involve 2 derivatives of Riemann (which is straightforward) or else expressions like like $\Gamma_{ijv,vv} \Gamma_{klu}$; after removing the two derivatives from this term, we obtain $\Gamma_{ijv} = -\Gamma_{vij} = -K_{ij(v)}$ and $\Gamma_{klu} = -\Gamma_{ukl} = -K_{kl(u)}$ in our gauge choice (where $\Gamma_{abc} = g_{cd} \Gamma^{d}_{ab}$).  We can ignore terms with derivatives acting on the area element because they cannot have weight more than 2.

The end result is an expression which involves at most two differentiations with respect to the Riemann tensor: one to get $H_{vv}$ and a second to identify the term with weight 3 or higher which gets differentiated twice with respect to $v$.  Recall that we can drop any term with two or more factors with positive Killing weight.

We use the standard convention where $\frac{\partial}{\partial R_{abcd}}$ has the same symmetries of the Riemann tensor, and is normalized so that $\delta X = \delta R_{abcd} \frac{\partial X}{\partial R_{abcd}}$.  This convention gives us the following symmetry factors when differentiating:
\begin{eqnarray}
\frac{\partial}{\partial R_{uvuv}},\, 
\frac{\partial}{\partial R_{uiuj}},\,
\frac{\partial}{\partial R_{vivj}},\,
\frac{\partial}{\partial R_{vijk}}: 4 \\
\frac{\partial}{\partial R_{uvui}},\, 
\frac{\partial}{\partial R_{vuvi}}: 8
\phantom{\frac{\partial}{\partial R_{vuvi}}}
\end{eqnarray}
from counting the number of equivalent ways to order the indices.  In addition, all terms have a factor of $1/2$ coming from differentiating Riemann with respect to the inverse metric using \ref{diffR}, a factor of 2 coming from the coefficient in Eq. (\ref{EOM}), a factor of $-2\pi$ coming from the coefficient in Eq. (\ref{2nd}), and some will have signs flipped due to manipulating $\Gamma$'s, either from taking the covariant derivative of a covector, or rearranging the order of indices of a $\Gamma$.

The first term in the entropy density $s$ comes from differentiating the Lagrangian with respect to $R_{uvuv}$.  This places a $\nabla_v \nabla_v$ outside the rest of the expression.  Removing these derivatives does \emph{not} give us $s_\text{Wald}$ due to the fact that $\nabla_v \ne \partial_v$.  However, for purposes of calculating the JKM piece of the entropy for $f(\text{Riemann})$, only the $\partial_v \partial_v$ piece contributes.  Thus we remove the factors of $\partial_v \partial_v$, and keep only those terms which are of the JKM form.  This gives us all terms coming from differentiating with respect to $R_{uvuv}$ (i.e. the Wald entropy density) \emph{except} for those terms which are fully boost-invariant (i.e. the Iyer-Wald entropy density):
\be
s^{(\text{JKM})}_1 = s_\text{Wald} - s_\text{IW}.
\ee
It is convenient to add the $s_\text{IW}$ term back in at this stage to obtain $s_\text{Wald}$.

The remaining JKM terms are:
\be
s^{(\text{JKM})}_2 = 32\pi \frac{\partial^2 L_g}{\partial R_{vkvl} \partial R_{uiuj}} \Gamma_{iju} \Gamma_{klv},
\ee
which comes from the $\Gamma^{v}_{ij} \partial_v$ term of $\nabla_i \nabla_j$,
\be
s^{(\text{JKM})}_3 = 
128\pi \frac{\partial^2 L_g}{\partial R_{vjkl} \partial R_{uvui}} \Gamma_{iju} \Gamma_{klv},
\ee
which comes from the terms in $\nabla_v \nabla_i R_{vjkl}$ containing $\Gamma^{v}_{ik}$ or $\Gamma^{v}_{il}$,
\be
s^{(\text{JKM})}_4 = 
-128\pi \frac{\partial^2 L_g}{\partial R_{vuvj} \partial R_{uvui}} \Gamma^k_{iu} \Gamma_{jkv},
\ee
where the $\Gamma^k_{iu}$ comes from $\nabla_i$ acting on the $u$ index of $R_{vuvj}$, and finally there is a derivative term
\be
64\pi \nabla_i \left( \frac{\partial^2 L_g}{\partial R_{vjvk} \partial R_{uvui}} \Gamma_{jkv} \right),
\ee
which arises if the $\nabla_i$ acts on a different Riemann curvature term than the $\nabla_v$ acts.  We can break this term up further using the formal expression $\nabla_i = D_i + K_i$, where $D_i$ is the derivative associated with parallel translating $i$-, $u$-, or $v$- indices \emph{along} the $D-2$ dimensional horizon slice, while $K_i$ is the connection corresponding to the extrinsic curvature, i.e. that part of $\Gamma_{ia}^b$ which contains exactly one $v$- or $u$- index.  The $D_i$ term is a total derivative:
\be
s^{(\text{JKM})}_5 = 64\pi D_i \left( \frac{\partial^2 L_g}{\partial R_{vjvk} \partial R_{uvui}} \Gamma_{jkv} \right),
\ee
which may be dropped when integrated along a compact horizon (or when evolving between slices of the horizon that differ only in a region with compact support).  The remaining pieces come from acting with $\Gamma_{ia}^{b}$ on the expression in parentheses (bearing in mind, when acting with the $\Gamma$'s that in the denominator reverses the role of downstairs and upstairs indices):
\begin{eqnarray}
s^{(\text{JKM})}_6 =
-64\pi \frac{\partial^2 L_g}{\partial R_{vjvk} \partial R_{ului}} \Gamma_{ilu} \Gamma_{jkv} 
\phantom{+ \frac{\partial^2 L_g}{\partial R_{vjlk}}}
\\ \nonumber
+128\pi \frac{\partial^2 L_g}{\partial R_{vjlk} \partial R_{uvui}} \Gamma_{ilu} \Gamma_{jkv}
+128\pi \frac{\partial^2 L_g}{\partial R_{vuvj} \partial R_{uvui}} \Gamma^k_{iu} \Gamma_{jkv}.
\end{eqnarray}
Adding up $s_\text{Wald} + s_2 + s_3 + s_4 + s_5 + s_6$, there are some cancellations and we obtain the total entropy
\be
S = \int d^{D-2}x\,\sqrt{g}\,(s_\text{Wald} - s_2 + s_5)
\ee
which matches Eq. (\ref{Dong}) after removing the total derivative term $s_5$.

Calculating the increasing entropy for actions which include derivatives of the Riemann tensor should be straightforward, if tedious.

\end{document}